\documentclass[amssymb,prl,aps,twocolumn,amsmath,showpacs]{revtex4-1} 
\usepackage{graphicx}
\usepackage{amsmath}
\usepackage{hyperref}

\begin{document}

\title{Droplet evolution in expanding flow of warm dense matter}
\author{Julien Armijo}
\email[Corresponding author: ]{julienarmijo@gmail.com}
\affiliation{Lawrence Berkeley National Laboratory, Berkeley, California, USA}
\author{John J. Barnard}
\affiliation{Lawrence Livermore National Laboratory, Livermore, California, USA}
\date{\today}

\begin{abstract} 

We propose a simple, self-consistent kinetic model for the evolution of a mixture of droplets and vapor expanding adiabatically in vacuum after rapid, almost isochoric heating. We study the evolution of the two-phase fluid at intermediate times between the molecular and the hydrodynamic scales, focusing on out-of-equilibrium and surface effects. We use the van der Waals equation of state as a test bed to implement our model and study the phenomenology of the upcoming second neutralized drift compression experiment (NDCX-II) at Lawrence Berkeley National Laboratory (LBNL) that uses ion beams for target heating.We find an approximate expression for the temperature difference between the droplets and the expanding gas and we check it with numerical calculations. The formula provides a useful criterion to distinguish the thermalized and nonthermalized regimes of expansion. In the thermalized case, the liquid fraction grows in a proportion that we estimate analytically, whereas, in case of too rapid expansion, a strict limit for the evaporation of droplets is derived. The range of experimental situations is discussed.
\end{abstract}

\pacs{64.70.fm, 51.10.+y}

\maketitle

\section{I. Introduction}

Warm dense matter (WDM) conditions, corresponding roughly to densites $0.01 < \rho/\rho_{\rm{solid}} < 10$ and temperatures $ 0.1 \rm{eV} < T < 10$ eV, can be defined as the region of thermodynamic space corresponding to the double crossover from degenerate to non-degenerate and from weakly to strongly coupled matter \cite{Lee03}, so that the ``easy" limiting descriptions in terms of cool plasma and hot condensed matter meet and have to be somehow connected to each other. This problem is drawing growing attention because of the serious theoretical challenges involved, and because of the occurence of WDM in the contexts of inertial fusion energy (IFE), astrophysics (planet cores), and laser ablation for materials processing, nanoparticles formation and film deposition \cite{Eliezer, Albert, AmorusoPRB}.


Generally, WDM experiments are inertially confined and explosive. Rapid heating of the material is required, so that the energy deposition (by lasers, ions, neutrons, electrical discharges, etc.) is faster than its release through hydrodynamic expansion. Pressures in the kbar to Mbar range can be reached before giving rise to supersonic expansion with typical outflow velocities of several km/s. 

The two-phase region of the phase diagram belongs only partly to the WDM regime, with its high temperature part near the critical point. However, any WDM experiment will almost inevitably lead to two-phase conditions during the expansion and cooling process.
This happens either from below the critical point (ion heating experiments at
Gesellschaft f\"ur Schwerionenforschung in Darmstadt \cite{TahirPRL, TahirNIMAB}, or second neutralized drift compression
experiment (NDCX-II) at Lawrence Berkeley National Laboratory (LBNL) \cite{Seidl}, low fluence laser ablation, Z machines), or from above it (IFE, high fluence laser ablation, upcoming ion heating machines). In the first case an overstretched liquid fragments and evaporates into a mixture of droplets and gas, whereas  in the second case a hot supersaturated gas nucleates  small clusters while expanding. In both cases the flow becomes a \textit{plume} of gas and condensed clusters (most often liquid, so the term "droplet" is appropriate) : the monophasic liquid or gas has undergone phase separation with the creation of \textit{surfaces} giving a non-trivial geometry to the fluid, which may \textit{a priori} affect its dynamical properties.

Recently, there has been significant progress in the observation of those two-phase flows, from the early ablation and plume expansion stages in the the ps and ns timescales \cite{Sokolowski-Tinten, Lindenberg, Zhang07, Oguri} to the late $\rm{\mu}$s timescale evolution including ÒpostmortemÓ
analysis of the clusters \cite{AmorusoSi, Lescoute}.

Basic questions arise when considering a two-phase flow. First, what is the droplets' size and distribution, and how do they evolve during the expansion? Second, are the droplets and the gas in thermal equilibrium? The answer can determine the conditions of validity for hydrodynamic approaches based on the Maxwell construction or any two-phase equation of state
(EoS) that assumes local equilibrium.

So far, two main approaches have been used. On one hand, molecular dynamics (MD) codes \cite{Holian, Ashurst, Toxvaerd, PerezPRL, PerezPRB, Lorazo, Upadhyay} compute the dynamics of each particle separately, and have given powerful insight into the processes of fragmentation, phase explosion, and the different mechanisms for ablation, but they are inherently limited to only treat the early times ($< 1$ns, 
\footnote{Typically, when simulating particles of mass $m$ interacting via a Lennard-Jones potential of range $\sigma$ and depth $\epsilon$, the natural molecular time scale is $\tau = (m \sigma^2/\epsilon)^{1/2} \sim100$fs, see e.g. \cite{PerezPRB}.}), and with a small number of particles ($\sim 10^{7}$). On the other hand, hydrodynamic codes \cite{Vidal, More, Barnard, Lescoute, Chimier09} can model experiments completely, but they deal with mesoscopic fluid cells, and often rely on crude approximations concerning the molecular and kinetic processes involved. Complex hydrodynamic codes including a treatment of the kinetics of phase change processes and surface effects in each cell are under development \cite{Lescoute, NIFEder}, but providing a complete and reliable description of a whole WDM experiment is still a challenge.

Better understanding of two-phase flows should be helpful for the preparation of experiments, including the diagnostics, and for the interpretation of the data. In IFE especially, the problem of debris dynamics is a crucial issue due to their impact on the optics and other components of the target chambers \cite{NIFEder, NIFKoniges}.
 
In this paper we propose an alternative approach to study two-phase flows in the cool or late time WDM situations. 
Our model was initially conceived to predict the phenomenology of the upcoming target heating experiments with the NDCX II machine at LBNL where an ion beam will almost isochorically heat a thin metallic foil to temperatures of about 1eV. However, the model should apply to any two-phase flow.

The core of our model is a self-consistent set of kinetic rate equations for the particle and energy fluxes between a droplet and the surrounding gas in an expanding Lagrangian cell. This
set of equations can be applied to any two-phase EoS. The
computing cell is considered as part of a larger hydrodynamic
code, but in this paper we only consider one cell containing
one droplet. We also neglect several features that could be
added in the near future.

To implement the kinetic equations and explore the patterns
of two-phase expansion, we use the van derWaals fluid model
that makes it possible to build a complete set of thermodynamic
functions.We thus demonstrate the ability of the kinetic model
to simulate nonequilibrium two-phase flows in the wide range
between themolecular and hydrodynamic scales. In particular,
we use it to distinguish the different regimes of two-phase
expansion: on one side, quasi- or fully thermalized; on the other
side, nonthemalized.We show that this distinction depends on
the initial target dimensions and the initial temperature. We
then study those regimes analytically and numerically.

\section{II. Background}
\subsection{A. Expanding two-phase flows. Supercritical and subcritical cases}

The model that we propose lies at a mesoscopic scale between the molecular and hydrodynamic scales, so we need some preliminar assumptions.
Our computing cell is considered as an elementary piece of a larger hydrodynamic code describing an expanding flow. The linear strain rate $\eta$ characterizes the expansion of the cell $L=L_0(1+\eta t)$, where $L_0$ is the initial cell size. We define the hydrodynamic time $t_{\rm{hydro}}=\eta^{-1}$. In the following, we assume rapid heating ($t_{\rm{heating}} < t_{\rm{hydro}}$) so that energy deposition in the material is almost isochoric. For simplicity, we assume instantaneous energy deposition. 

To get insight into the global flow, it is interesting to review some analytical and numerical results.
The self similar rarefaction wave (SSRW) is the solution \cite{Landau} describing the one-dimensional (1D) expansion of a perfect gas (semi-infinite at $z<0$) of adiabatic coeficient $\gamma$ after instantaneous uniform heating at the initial temperature $T_0$. Denoting $c^s_0$ the sound speed  in the fluid at $T_0$, the SSRW solution describes an outward expanding front travelling at the outflow velocity $v_0=2 c^s_0/(\gamma -1)$, which is $3 c^s_0$ for a perfect monoatomic gas, while the inward rarefaction wave propagates at $c^s_0$ \cite{Barnard}. Note that the SSRW can be computed semi-numerically for any EoS of a non-ideal gas \cite{Anisimov99} and has been validated by MD simulations \cite{PerezPRB}.

\begin{figure}[htbp]
\begin{center}
\includegraphics[width=8.5 cm]{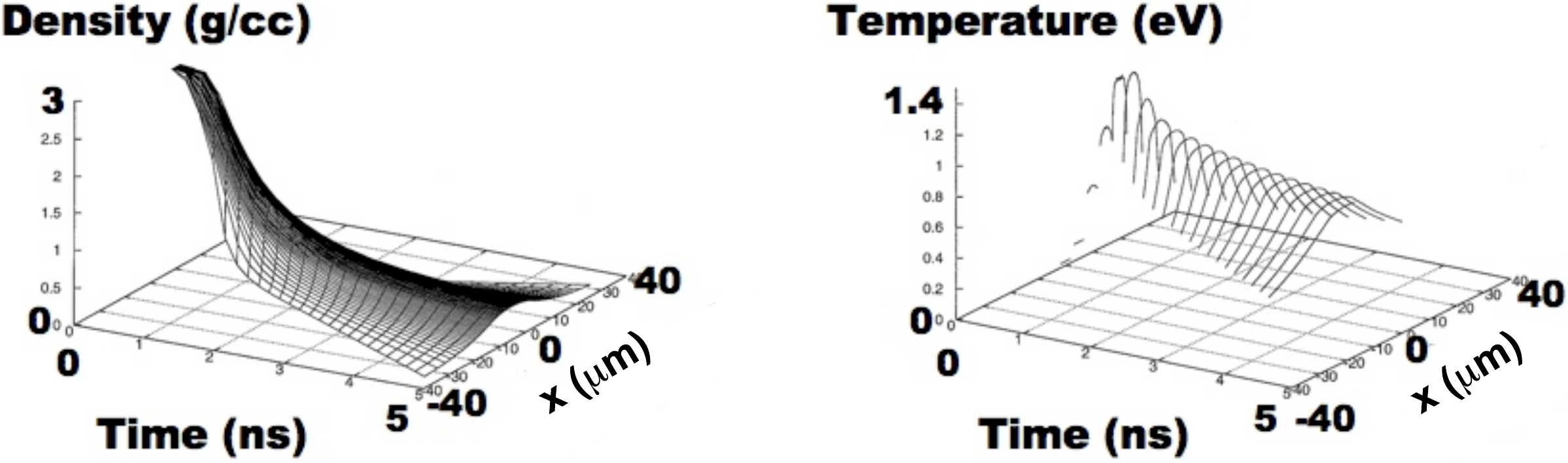}
\caption{Hydrodynamic calculation with DPC code of NDCX II reference case, from \cite{Barnard}. A 3.5$\mu$m-thick Al foil is heated within 1ns with an ion beam and subsequently cools down during adiabatic expansion}
\label{fig.DPC}
\end{center}
\end{figure}

As an example of a numerical simulation of expanding flows, Fig.~\ref{fig.DPC} shows a hydrodynamic calculation with the code DPC using an EoS based on Maxwell construction \cite{More}. Here the liquid and gas are assumed in equilibrium, which is not kinetically justified (see Sec. IV.B), and the outflow velocity is about $8$km/s after $10$ns.

In the following we assume a flow with linear speed profile and outflow velocity $v_0 = 3 c^s_0$, but it is worth remarking that this is quite simplistic. In particular, several numerical works \cite{Vidal, Barnard, Chimier09, Anisimov99} have reported the occurrence of "plateaus", that is, zones of nearly constant density, related to the fluid zones entering into the two-phase regime.

\begin{figure}[htbp]
\begin{center}
\includegraphics[width=8.5 cm]{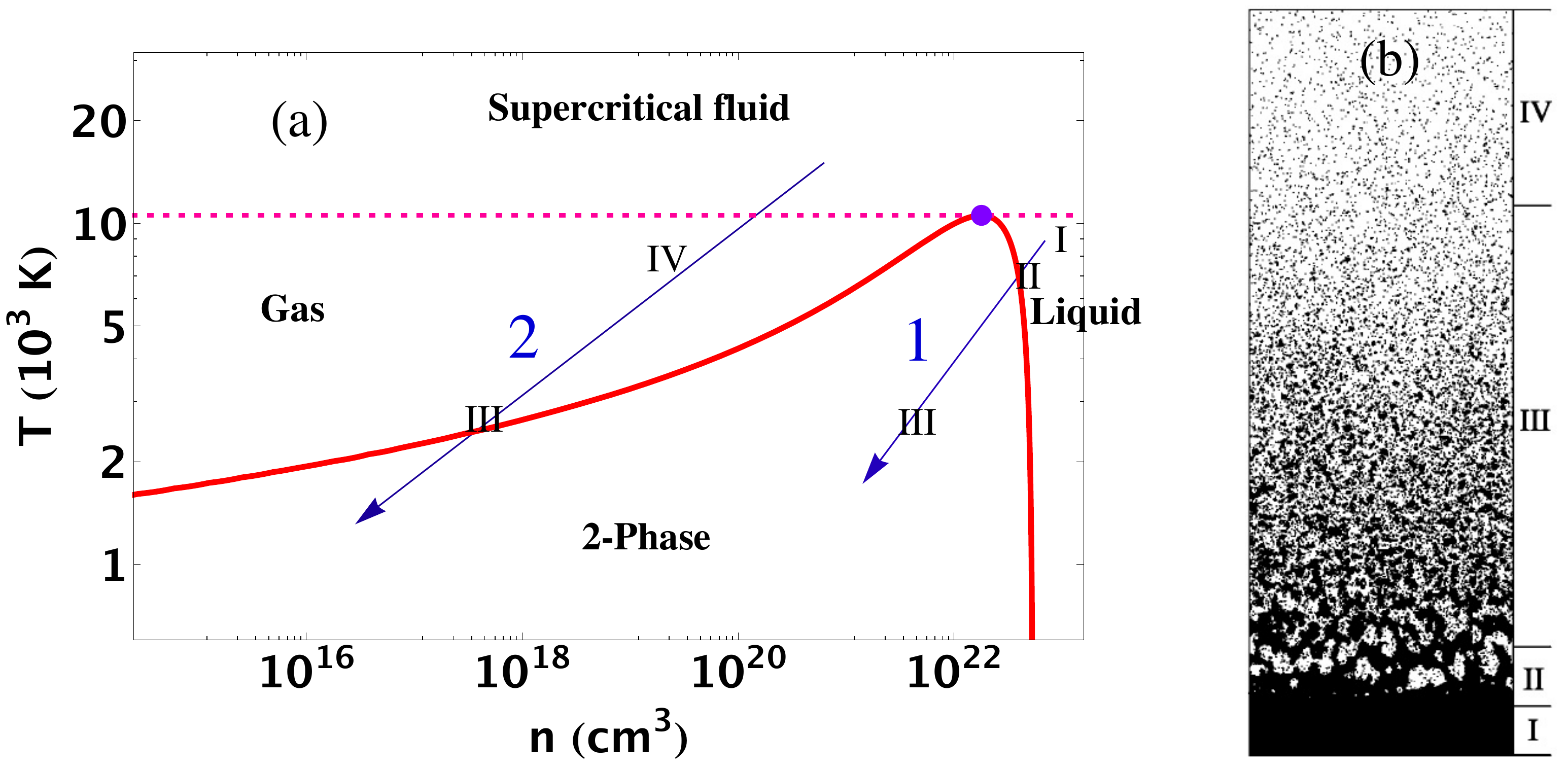}
\caption{(Color online)  (a) Phase diagram of the van der Waals EoS for Al (see Sec. III) showing the liquid and gas binodals (solid lines), and a schematic representation of the subcritical (arrow 1) and supercritical (arrow 2) cases of two-phase expansion. (b) 2D MD simulation of laser ablation with inhomogeneous initial temperature, from \cite{PerezPRL}, showing material in various situations of two-phase expansion, which we also locate qualitatively on the schematic classification of Fig.~\ref{fig.regimes}.a (roman numbers).}
\label{fig.regimes}
\end{center}
\end{figure}

Let us now present our simple classification of two-phase expanding flows, which is based on \textit{equilibrium} thermodynamics, and is similar to the one in \cite{Cheng}. Of course, such procedure cannot account for the complexity of non-equilibrium situations encountered in experiments or simulations \cite{Lindenberg,Oguri,PerezPRL, PerezPRB, Lorazo, Cheng}. 
The two-phase regime exists only for temperatures $T<T_c$, and for an average density between the value at the liquid and gas binodals, as seen in Fig.~\ref{fig.regimes}.a. 
Thus, a piece of fluid expanding near thermodynamic equilibrium can enter the two-phase regime, which amounts to partially undergoing the liquid-gas first-order phase transition, in only two ways: by crossing the liquid binodal, which we call the \textit{subcritical case}, or by crossing the gas binodal, which we call the \textit{supercritical case}.

In the subcritical case, which corresponds to the calculation of Fig.~\ref{fig.DPC} when it is mapped onto the corresponding phase diagram, an expanding monophasic liquid reaches the liquid binodal and becomes overstretched. The equilibrium configuration for a piece of fluid in the two-phase region is a mixture of liquid droplets (of yet undetermined size) and gas whose densities are at the binodals for the same temperature. We call \textit{fragmentation} the transformation from the monophasic liquid to the non-connected cloud of droplets.
In the supercritical case, achieved if the material is initially heated to higher temperatures, a monophasic gas becomes supersaturated after crossing the gas binodal, and we call \textit{nucleation} the process by which a certain distribution of liquid droplets is created.

Figure \ref{fig.regimes} represents the two cases and the various experimental situations that they involve. In Fig.~\ref{fig.regimes}.a, we show the Van der Waals phase diagram for Al that we use in the following, and a schematical representation of the sub- and supercritical cases of two-phase expansion (arrow 1 and 2). In Fig.~\ref{fig.regimes}.b, reproduced from \cite{PerezPRL}, one sees the particle distribution in a 2D MD simulation of laser ablation. 
Due to the inhomogeneous energy deposition, different types of thermodynamic evolutions are seen at a same time, and we use this picture to illustrate the various situations of our classification, although not following exactly the terminology of the original work
\footnote{The classification that we use through this paper is not fully compatible with the one in \cite{PerezPRL, PerezPRB, Lorazo}, where the notion of ``fragmentation'' is used in supercritical conditions. Such use is absent in our perspective since the concept of fragmentation of a liquid in several pieces separated by gas is associated with the first-order liquid-gas phase transition, which only exist for $T<T_c$. There, the latent heat, and the surface tension, that allows spatial separation of the two phases, are defined.
In the works \cite{PerezPRL, PerezPRB, Lorazo}, dense \textit{vs} dilute parts are distinguished from a continuous supercritical medium with density fluctuations, but the criterium used is not related with the phase transition upon which our classification is based.
}.
In zone I, the dense material is still continuous. In zone II, the expanding liquid has undergone cavitation of gas bubbles. In the upper zone II and in zone III, the liquid is fragmented and the material has entered the two-phase regime in the subcritical case. In zone IV, 
the fully atomized, expanding material is likely to reach the gas binodal in a later stage, thus corresponding to the supercritical case. In the upper zone III, small clusters are present, which could originate from fragmentation at high temperature (subcritical case) or recent nucleation (supercritical case). This is why in Fig.~\ref{fig.regimes}.a., where the different zones are placed \textit{qualitatively}, two positions are proposed for zone III.

\subsection{B. Initial droplet size}

Both cases lead to droplets formation. In order to initialize the kinetic model that we present further, it is necessary to know the initial droplet size at the onset of the two-phase regime.

In the subcritical case, the overstretched liquid starts cavitating (see Fig.~\ref{fig.regimes}.b, zone II), which we call the \textit{bubbles regime} and then the bubbles percolate until the liquid phase is not continuous anymore (see Fig.~\ref{fig.regimes}.b, zone III), which we call the \textit{droplets regime}. We assume that the droplets regime starts when the gas and liquid volumes are equal: $V_g = V_l$, which is justified by an argument of surface energy minimization.

The mean droplet's size in a fragmentation scenario can be obtained by considering a balance between the disruptive inertial forces and the restoring surface tension \cite{Holian}. The model proposed initially by Grady \cite{Grady} has been abundantly validated by MD calculations \cite{Holian, PerezPRL, Upadhyay} in 2D and 3D, and is in very good agreement with measurements on He jets \cite{Knuth}. We note that the scaling of the mean radius $R$ of the droplet can be obtained by just setting to unity the Weber number $We\equiv \rho R v^2 /\sigma$ \cite{pilch87}, where $\sigma$ is the surface tension $\rho$ the liquid mass density, and $v=\eta R$ the typical velocity difference across a piece of fluid of size $R$. $We$ is the ratio of the surface energy to the inertial energy. In any dimension this criterion yields
\begin{equation}
\rm{We} \ \sim \  1 \qquad  \Rightarrow \qquad   R  \ \sim \  \left( \frac{\sigma}{\rho \eta^{2}}\right)^{\frac{1}{3}}.
\label{eq.req}
\end{equation}
Several values of order 1 have been proposed for the prefactor in this scaling law, either from analytical estimates (prefactor $15^{1/3}=2.47$ in \cite{Chimier09}), or from fits to MD simulation results.
In \cite{Ashurst}, it was shown that both 3D MD results with a homogeneous strain rate $\eta$ and data from helium free jets experiments from \cite{Knuth} could be fitted to Eq.~\ref{eq.req} with the same prefactor, thus validating this law over almost 8 orders of magnitude in cluster mass (the experimental fragments cover larger sizes than the numerical ones).

Concerning the size distribution of droplets resulting from fragmentation, MD simulations have shown clearly that it is essentially exponential \cite{Holian,Upadhyay, Toxvaerd, Ashurst}, which is consistent with simple models of entropy maximization \cite{Holian}.

By contrast, it is not so clear how to describe the initial situation in the supercritical case. This task requires one to choose a model for nucleation, or to input results from MD calculations. Nucleation of clusters from a supersaturated vapor is the situation of nucleation whose kinetics is the easiest to model theoretically \cite{Balibar}, but still choices have to be made \cite{Lescoute}, that are beyond the scope of this paper. Any model for nucleation will depend crucially on surface tension, so we make the remark here that estimating the surface tension for small droplets is delicate because of its enhancement at small sizes \cite{Moody, Tolman}.

\section{III. Model}

\subsection{A. Van der Waals fluid model}

The kinetic model that we present in Sec. III.C is applicable to any EoS. In this paper, for a qualitative investigation of the two-phase expansion regimes, including kinetic and surface effects, and with emphasis on the late time and low temperature limits, we use a van der Waals (VdW) fluid model, which for convenience we describe first.
It is important to note, however, that the VdW model is not intended to provide a highly accurate description of a fluid, especially in WDM or supercritical conditions where ionization and radiation effects can be strong.

With only two parameters, the VdW EoS is the simplest EoS describing the coexistence of a liquid and a gas phase, and has already been used for theoretical studies of dynamic two-phase processes \cite{OnukiPRL, OnukiPRE}. 
All the thermodynamic functions can be derived from the expression for the mean-field potential energy per particle in such fluid: $U= +\infty$ if $n>1/b$ and $U=- a n$ if $n < 1/b$, where $n$ is the particle density, $b$ stands for the incompressible volume of the particles, and $a$ represents the mean-field attractive energy between them. 

The bulk VdW energy of $N$ particles at temperature $T$ is $E=N (c_{v} T - a n)$.
It can be shown that the specific heat $c_{v}$ is independent of $n$ and can only depend on $T$ \cite{Reif}, so that one has to choose necessarily $c_{v} = \frac{3}{2}k_{B}$, where $k_B$ is the Boltzmann constant, if one wants the EoS to match the perfect monoatomic gas in the dilute limit. Writing the partition function, one obtains the other thermodynamic functions. In particular, the pressure is $P=k_{B}T/(v-b) - a/v^2$ where $v=1/n$ is the volume per particle. This expression implies that the isobars (isotherms) are a cubic relationship between $T$ and $v$ ($P$ and $v$). Hence, below a certain critical temperature $T_{c}$, an unstable zone of negative compressibility appears in the phase diagram, limited by the two spinodals. We obtain the  equilibrium density of the two stable phases that can coexist at certain $(P,\ T)$ by numerically performing the Maxwell construction, which consists of solving $P_l=P_g$ (i) and $\mu_l=\mu_g$ (ii) simultaneously, where $\mu$ denotes the chemical potential and the subscripts $l$ and $g$ stand for $liquid$ and $gas$, respectively, all throughout this paper. (ii) is equivalent to $ \int_l^{g}v dP = 0$ and thus $\int_l^{g}{P(v) dv} = P_{l,g} (v_g - v_l)$ \cite{Reif}.

Introducing the reduced temperature $\theta =k_{B}T/l_0$, where $l_0=a/b$ is the latent heat at $T=0$, and two dimensionless parameters that are small in the low temperature limit: $v_g=b/\delta$ and $v_l = b (1 +\gamma)$, equations (i) and (ii) become: 
\begin{align}
\frac{\theta}{\gamma}-\frac{1}{(1+\gamma)^{2}} & = \frac{\theta}{\frac{1}{\delta}-1} -\delta^{2}
\label{eq.Max1} \qquad \rm{and} \\
\theta \rm{ln}\left( \frac{ \frac{1}{\delta} -1}{\gamma}\right) + \delta -\frac{1}{1+\gamma}
& =\bigg(\frac{\theta}{\frac{1}{\delta} -1} -\delta^{2}\bigg) \bigg(\frac{1}{\delta} -(1+\gamma)\bigg).
\label{eq.Max2}
\end{align}
It is worth remarking that $T_{c}=8 a/27 b$, so $\theta_c =8/27 \simeq 0.3$, and therefore one expects that calculations in the "low temperature limit" ($\theta \ll 1$) should be a good approximation as soon as one is not considering the vicinity of the critical point.

\begin{figure}[htbp]
\begin{center}
\includegraphics[width=8.5 cm]{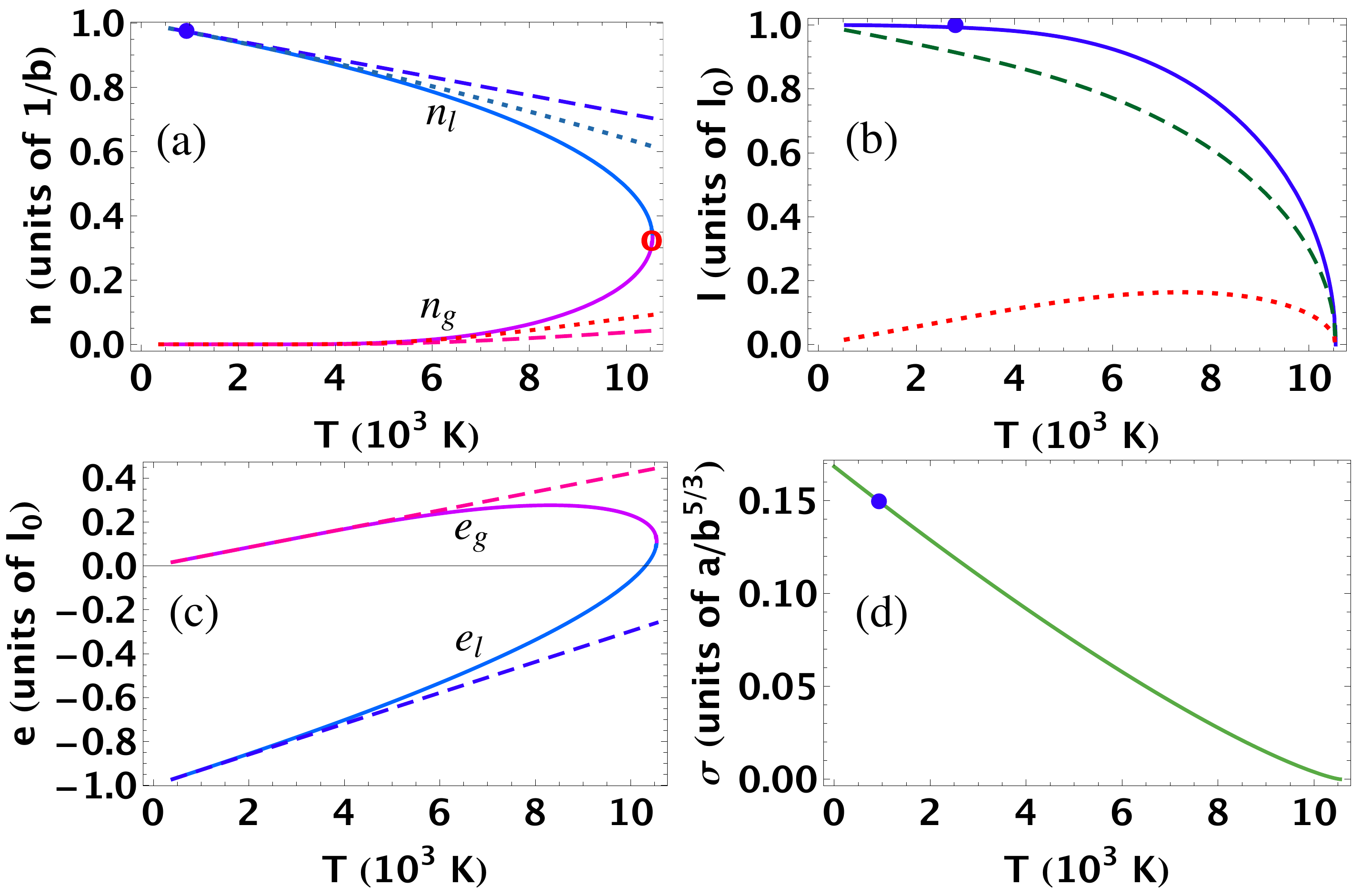} 
\caption{(Color online) Van der Waals thermodynamic functions for Al, in VdW units. 
The filled circles represent experimental data and the empty cirlce the critical point obtained from the fits.
 (a) Liquid and gas densities (solid lines) with first (dashed lines) and second order (dotted lines) low $T$ approximations. (b) Latent heat (solid lines) decomposed in the first (dashed lines) and second (dotted lines) term of Eq.~\ref{eq.latHeat}. (c) Bulk energies per particle, with first order low $T$ approximations Eq.~\ref{eq.energies} (dashed lines). (d) Surface tension (Eq.~\ref{eq.sigma}).}
\label{fig.EoS}
\end{center}
\end{figure}

Figure \ref{fig.EoS} gathers the thermodynamic functions of our VdW model for aluminum.
Figure~\ref{fig.EoS}.a shows the numerical result of the dimensionless Maxwell construction where the VdW parameters $a=9.1\times10^{-35} \rm{erg.cm^{3}}$ and $b=1.85\times 10^{-23} \rm{cm^3}$, giving $l_0 =3.07 eV$, have been adjusted to fit this material. For that, we impose that the VdW liquid density matches the aluminum liquid density $n_l(T_m)= 5.26 \times 10^{22} $cm$^{-3}$ at the melting point $T_{m} = 933.5$K ($=0.026 l_0$) \cite{CRC} and that the VdW latent heat (shown in Fig.~\ref{fig.EoS}.b)
\begin{equation}
 l= a (n_l-n_g) + P_{l,g}\bigg(\frac{1}{n_l} - \frac{1}{n_g}\bigg) 
 \label{eq.latHeat}
 \end{equation}
coincides with the experimental value $l(T_b)=4.88 \times 10^{-12}$erg/atom for aluminum at the boiling temperature $T_{b}=2792$K ($=0.078 l_0$) \cite{CRC}. Note that the critical parameters obtained in this way are consistent with the best estimates to date, although not very precisely \footnote{We obtain a critical temperature $T_{c}=10545$K and a critical density $\rho_{c}=0.81$g/cc whereas in [I. Lomonosov, Laser Part. Beams 25, 567 (2007)], a comprehensive synthesis of experimental data and models leads to the estimations $T_{c}=6250$K and $\rho _{c}=0.70$g/cc for aluminum.}.

In Fig.~\ref{fig.EoS}.a are also displayed the simple, useful approximations for $n_l$ and $n_g$ at lowest orders in $\theta$ that one obtains directly from Eq.~\ref{eq.Max1} and \ref{eq.Max2} :
\begin{align}
n_l &\simeq \frac{1}{b} (1 - \theta - \theta^{2}) \label{eq.nl}  ,\\
n_g & \simeq \frac{1}{b \theta} \exp\bigg(-\frac{1}{\theta (1+\theta)}\bigg)  .
\label{eq.ng}
\end{align} 
Note that Eq.~\ref{eq.ng} is, at lowest order in $\theta$, equivalent to the Clausius-Clapeyron formula applied to a perfect gas. We also show in fig~\ref{fig.EoS}.c the approximations at first order in $\theta$ for the liquid and gas bulk energies per particle 
\begin{equation}
e_g\simeq \frac{3}{2} k_{B} T      \quad , \quad   e_l\simeq \frac{5}{2} k_{B} T -\frac{a}{b}.
\label{eq.energies}
\end{equation}

For the surface tension, Van der Waals himself had already proposed to model it using density gradients \cite{OnukiPRE}, but we have chosen to use a simple formula that is universally verified in simple fluids \cite{Buff}
\begin{equation}
 \sigma \propto (1-T/T_{c})^{1+r} \quad    \rm{with} \quad   r=0.27.
\label{eq.sigma}
\end{equation} 
To model aluminum, we fit this formula to the experimental value $\sigma (T_m) = 1050 \rm{erg/cm^2}$ \cite{Sarou-Kanian03}, as shown in fig.~\ref{fig.EoS}.d. Note that in the following, the total liquid energy in the cell is $E_l = N_l e_l + \sigma S_l$, where $S_l$ is the surface area of the droplet.

\subsection{B. Kinetic equations. Validity condition}

\begin{figure}[htbp]
\begin{center}
\includegraphics[width=8.5cm]{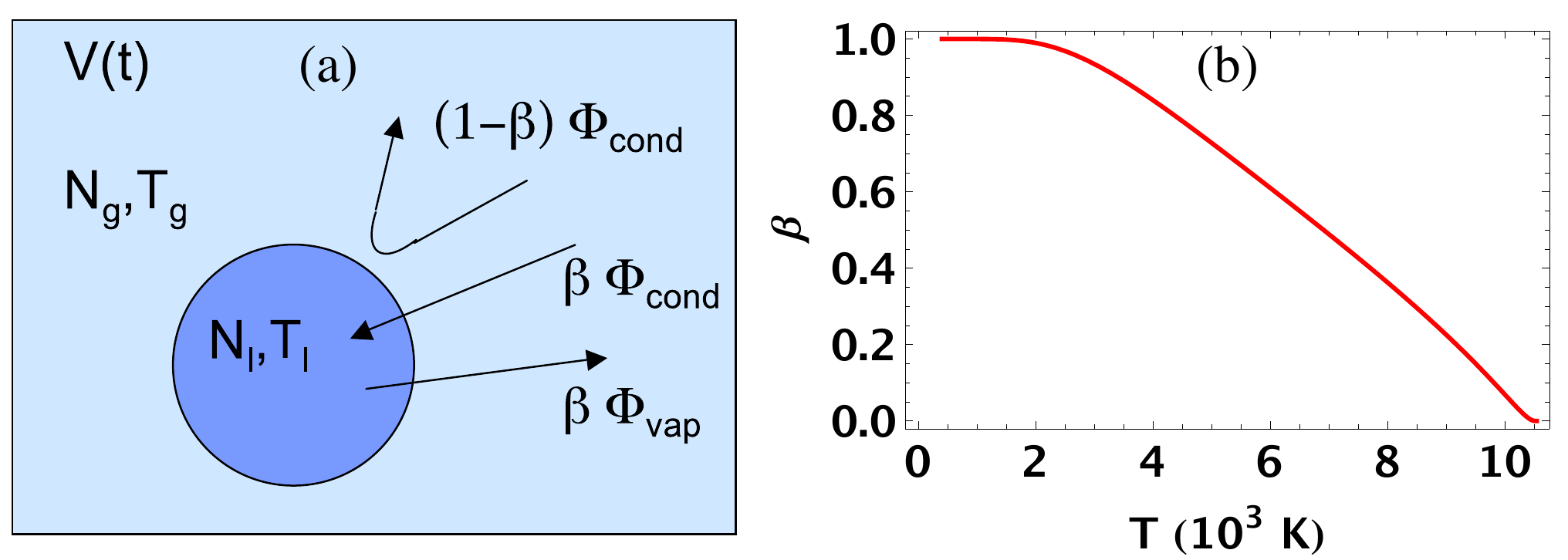}
\caption{(Color online)  The "droplet-in-cell" kinetic model. a) Sketch of the kinetic fluxes. b) Sticking coefficient $\beta$ calculated with formula from \cite{Nagayama} and our aluminum VdW parameters.}
\label{fig.model}
\end{center}
\end{figure}

Our goal is to compute the evolution of droplets in cells. In this paper we limit ourselves to the case of one droplet in one Lagrangian cell undergoing adiabatic expansion. It is a closed system out of equlibrium, and its complete description requires the determination of the four variables $N_l, n_l, T_l,T_g$. To compute their evolution, we need four rate equations: a liquid-gas particle exchange rate, an energy exchange rate, a total energy loss rate (work to the outside), and an internal equilibrium condition to determine the liquid density. As shown in fig.~\ref{fig.model}.a, the particle fluxes between liquid and gas are divided between evaporating, condensing, and condensing but not-sticking particles.

The volume expansion $V(t)$ shall later be prescribed by a global hydrodynamic code. For our study, we assume cylindrical symmetry and we use a simple model behavior \cite{Toxvaerd, Ashurst}
\begin{equation} 
 V(t) = V_0 (1+\eta_z t)(1+ \eta_r t)^{2}, 
 \label{eq.V(t)}
 \end{equation} 
where $V_0 = L_0^3$ is the inital cell volume and $\eta_z = (dv_z / dz)_{t=0}$ and $\eta_r = (d v_r/dr)_{t=0}$ are the axial and radial strain rates, respectively defined as the initial velocity gradients in the beam direction and the target plane.

The evaporation and condensation fluxes are computed using the standard Hertz-Knudsen formulas \cite{Hertz, Knudsen, Cercignani}
\begin{equation}
\Phi_{\rm{cond}}=n_g \sqrt{\frac{k_{B} T_g}{2\pi m}}  \ \  , \ \    \Phi_{\rm{vap}}=n_g^{*}(T_l, R) \sqrt{\frac{k_{B} T_l}{2\pi m}},
\label{eq.fluxes}
\end{equation}
where $m$ is the particle mass and $n_g^{*}(T_l, R)$ is the \textit{equilibrium gas density} for a droplet at temperature $T_l$ and of radius $R$. To estimate $n_g^{*}$, we use the Kelvin equation, which describes the increase of the equilibrium vapor pressure surrounding a droplet due to surface tension
\begin{equation}
 n_g^{*}(R) = n_g^{*}(\infty) \exp\bigg(\frac{2\sigma}{k_{B}Tn_l R}\bigg).
 \end{equation}
The Kelvin equation is approximate because its derivation assumes a perfect gas. Also, we use a constant value for $\sigma$, thus neglecting its increase at small radii \cite{Moody, Tolman}. Still, this approach is probably not too bad after all \cite{Powles}, and satisfactory enough for our qualitative purpose.

Considering mass conservation, and combining the two fluxes of Eq.~\ref{eq.fluxes}, the particle exchange rate equations are
\begin{equation}
\frac{d (N_l+N_g)}{dt}=0 \  \ , \  \
\frac{dN_l}{dt}= \beta(- \Phi_{\rm{vap}}+\Phi_{\rm{cond}}) S_l,
\label{eq.evap}
\end{equation}
where $S_l$ is the surface area of the droplet and $0 < \beta < 1$ the \textit{sticking coefficient} that is usually assumed of order 0.5. A recent study \cite{Nagayama} has proposed a simple expression for $\beta$ that is in good agreement with MD calculations for several simple fluids. This expression depends only on the ratio of the molecular volumes in the liquid and vapor phase: $\beta =(1-({v_l/v_g})^{1/3}) \exp(-\frac{1}{2} \frac{1}{({v_l/v_g)^{1/3}} -1})$ that we plot for our VdW model for Al in Fig.~\ref{fig.model}.b. Note however that, for the calculations presented in this paper, we use a constant value $\beta =0.5$.

Concerning the energy fluxes, the first equation comes from the assumption of adiabatic expansion of the cell : 
\begin{equation}
\frac{d(E_l +E_g)}{dt}=-P_g \frac{dV}{dt}.
\label{eq.dEtot}
\end{equation}
In this global energy loss rate we have neglected three terms that could be added in the near future. The first one is heat conduction between cells. This term may play a role, but it cannot be very important as we are considering a supersonic flow (see Sec. I).
The second neglected term is radiation. Radiation becomes indeed the dominant cooling mechanism at long times, as we see later, but it is negligible for the initial dynamics, so the approximation is reasonable, because our purpose in this paper is to study the expansion in a time range where the two phases are interacting and the system is not just a collection of isolated clusters flying in vacuum. The third neglected term is thermionic emission. One expects electrons to be thermally emitted from the droplet, taking away some energy. Non-neutral effects are totally absent from our model, but we expect that the associated cooling rates will be small compared to the adiabatic and evaporative cooling rates
\footnote{Using Richardson law, one can estimate that the rate of electron emission may be high at temperatures close to $T_c$ in the first ns of expansion. In the case where the electrons just fly away with no collisions with other droplets, the energy loss for the cell is expected to be smaller than the total work term in the early times and smaller than the radiative term at long times. In addition the build-up of a charge of the droplet should quickly limit the emission. In the opposite case where the emitted electrons would be recaptured by other droplets, the thermionic emission and recapture terms should only amount to an increase of the thermalization rate in Eq.~\ref{eq.dEl}, without causing a fundamental change to the picture.}.
 
The energy exchange rate between the liquid and the gas has contributions from the three fluxes of Fig.~\ref{fig.model}.a. The contribution of the colliding but non-sticking particles can be described with a flux proportional to the temperature difference $T_g -T_l$, with a relaxation coeficient $0<\alpha <1$ (see e.g. \cite{Bond} for more discussion). For the condensing gas particles, we make the simplest assumption, that each of them brings into the liquid the average gas energy per particle $e_g$.
For the evaporating particles, we assume that the energy they individually take away from the liquid depends only on the liquid state. We note it $e_g^{*}$ and define it as the energy of a virtual gas particle that would be in equilibrium with the droplet of radius $R$ at temperature $T_l$. 
For the VdW fluid, those energies are $e_g=c_{v} T_g - a n_g$ and $e_g^{*}=c_{v} T_l - a n_g^{*}(T_l, R)$, respectively.
Note that our definition of $e_g^*$ is totally analogous to the Hertz-Knudsen derivation of the mass evaporation rate of Eq.~\ref{eq.evap}.  We finally get the exchange rate
\begin{align}
\notag & \frac{dE_l}{dt}=  \big[ \beta  (- e_g^{*}  \Phi_{\rm{vap}}+ e_g  \Phi_{\rm{cond}}) \\ 
 & \qquad  \qquad   \ \ \  + (1-\beta) \alpha c_{v} (T_g-T_l) \Phi_{\rm{cond}} \big] \ S_l .
 \label{eq.dEl}
\end{align}

 Our set of kinetic equations is fully consistent in the sense that at equilibrium both mass and energy fluxes between the droplet and the gas are in balance. In particular, the average energy that a liquid particle takes from the rest of the liquid to evaporate is $e_g^* - e_l$. This term, which for the VdW fluid is $- a (n_g^* - n_l)$, corresponds exactly to the latent heat per particle for an arbitrary fluid $l = (e_g -e_l) - P_{l,g}(v_g-v_l)$, without  the second (work) term, which is expected since the latent heat is an enthalpy and we are here dealing with energy exchanges at constant volume.
 
 To our knowledge, our set of rate equations is an original model for the exchanges between a droplet and its vapor. Other sets of kinetic equations can be found for analogous systems (see e.g. \cite{Kryukov, Sazhin}), but they do not correspond to the purely kinetic regime that we are considering, because they deal with larger droplets ($R > 1 \mu$m) and longer time scales, more relevant to the fields of combustion or atmospheric sciences, so they need to combine the kinetic approach with the more classic hydrodynamic theories of droplet evaporation \cite{Fuchs}.
 
Our model for WDM situations is simpler, because we do not distinguish in the gas a Knudsen layer \textit{vs} a hydrodynamic layer. We assume that our computing cells are small enough that the gas density inside them is constant. The variations over the whole flow shall instead be treated by the global hydrodynamic code that determines the expansion of each kinetic cell. The kinetic and phase change processes in our description are driven by the hydrodynamic expansion, therefore, the validity condition of our model is  that the initial cell size should be much smaller than the initial dimensions of the expanding material :  
\begin{equation}
L_0 \ll \delta r, \delta z ,
\label{eq.valcond}
\end{equation}
where $\delta z$ is the initial foil thickness and $\delta r$ the heating beam diameter.
If Eq.~\ref{eq.valcond} is verified, the global hydrodynamic treatment is correct, with gradients properly resolved. Eq.~\ref{eq.valcond} is verified in the standard situations we consider, but can break down if the initial droplet or cell size is not small enough compared to the sample size.

\subsection{C. Equilibrium condition between droplet and gas} 

In order to get a closed system of equations for particle and energy fluxes, we still need one assumption. Our model is \textit{a priori} out of thermal equilibrium ($T_l \not=  T_g$), so the density of the liquid is not determined yet. It seems reasonable to assume pressure equilibrium between the droplet and the gas \cite{Lescoute}, because we expect that a few collisions are sufficient for the droplet to "experience" the gas pressure, and adjusting the liquid pressure to it requires only a small density change, due to the very low compressibility of the liquid. 

Due to the droplet curvature, the pressure equilibrium condition is the Laplace equation
\begin{align}
P_l -P_g =\frac{2 \sigma}{R}.
\label{eq.Laplace}
\end{align}
An exact numerical implementation of Eq.~\ref{eq.Laplace} is difficult, because it requires to solve a non-linear system at each timestep in order to determine the liquid density given a certain set of values \{V, $N_l$, $N_g$, $E_l$, $E_g$\}.

To simplify the condition, one can approximate the liquid density by the equilibrium value $n_l^*(\infty)$ for a flat interface ($R \rightarrow \infty$), but this is wrong for two reasons : first, because due to the fast expansion, the gas pressure is lower than the saturation value corresponding to the liquid temperature, and second, because of the Laplace compression term of Eq.~\ref{eq.Laplace}.
Running our model, we have checked that this raw approximation leads to important inaccuracies in the calculation of the pressure, especially at low temperatures where the Laplace term becomes dominant. Still, these errors do not cause important discrepancies in the global description of the droplet evolution, which we attribute again to the low compressibility of the liquid.

For more accuracy, we choose to compute the liquid density in perturbation from the flat equilibrium value :
\begin{align}
 n_l=n_l^*(\infty) \bigg(1+ \frac{\Delta P}{K_l(T_l)}\bigg),
 \label{eq.Peq}
\end{align}
where $K_l(T_l) =  n_l (\partial P / \partial n_l)_{T_l}$ is the isothermal bulk modulus of the liquid that we compute directly from the VdW EoS, and $\Delta P = 2 \sigma/R -(P_l(n_l^*(\infty),T_l) -P_g)$ is the pressure correction that we compute using Eq.~\ref{eq.Laplace}.
As we show in the next section, this perturbative approach of the pressure equilibrium condition is very satisfactory.

With Eq.~\ref{eq.V(t)}-\ref{eq.Peq}, we have a complete kinetic model for the evolution of a droplet and its vapor in a cell. 
In the following, we combine it with the VdW EoS to study the different regimes of two-phase expansion.

\section{IV. Results}

\subsection{A. NDCX II reference case (subcritical case)}

The reference case envisioned as an upcoming experiment on the NDCX II machine at LBNL consists of heating an aluminum foil of thickness $\delta z = 3.5 \mu$m with a short pulse of ions of duration $t_{\rm{heating}} \lesssim 1$ns, which makes the picture of rapid heating roughly correct \cite{Seidl}.
The beam profile is taken as a uniform disk of diameter $\delta r =1$mm. Initial temperatures up to 1eV are predicted for the expected beam fluences \cite{Barnard}.

\begin{figure}[htbp]
\begin{center}
\includegraphics[width=8.5 cm]{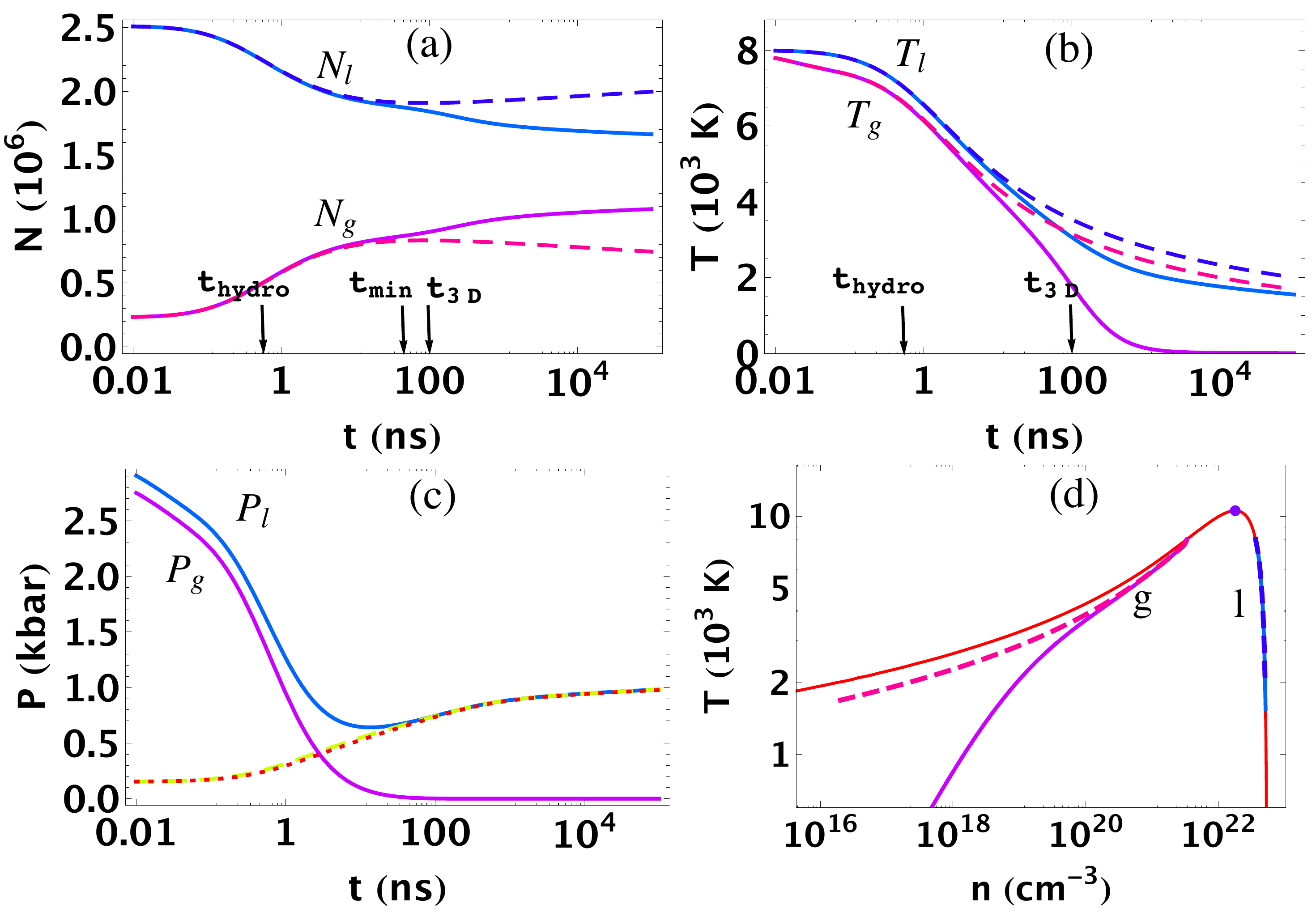}
\caption{(Color online)  Droplet and gas evolution in the NDCX II reference case. Initially the droplet of radius $R_0 = 25.4$nm and the gas have equal volumes and are in equilibrium at $T_0=8000$K. All liquiwd (gas) quantities are labeled with $l$ ($g$). Plotted are the time evolution of (a) the particle numbers and (b) the temperatures,  for a 1D (dashed lines) or a full 3D expansion (solid lines). (c) Pressure evolution in the 3D case. The pressure difference computed (dashed lines) and expected from Eq.~\ref{eq.Laplace} (dotted lines) are indistinguishable. (d) Trajectories in the phase diagram for the 1D (dashed lines) and 3D (solid lines) cases}
\label{fig.refcase}
\end{center}
\end{figure}

In Fig.~\ref{fig.refcase}, we present the numerical output of the model for a cell containing a droplet and gas initially at equilibrium at $T_0 = 8000$K with $V_l = V_g = V/2$, because this corresponds to the onset of the "droplets regime" (see Sec. II.B). As we mentioned previously, we make the crude assumption of a flow with linear speed profile and outward expanding speed $v_0=3c^s_0 \simeq 5.0$km/s on both sides $z>0$ and $z<0$, where the initial sound speed $c^s_0$ is estimated roughly as the thermal velocity $v_{\rm{th}}(T_0)=\sqrt{k_B T_0/m}$. Then, the strain rates in Eq.~\ref{eq.V(t)} are simply $\eta_z = 6 c^s_0 / \delta z$ and $\eta_r =6 c^s_0 / \delta r$. We display a full 3D case (solid lines) and a 1D case (dashed lines) where $\eta_r =0$. Here the hydrodynamic time is $t_{\rm{hydro}} =1/\eta_z \simeq 0.37$ns. The time $t_{3D}=1/\eta_r \simeq 107$ns can be considered as the time of the crossover to the 3D regime of expansion. The calculation is carried out with $\alpha = \beta = 0.5$ and we use the variable $u=\ln (t)$ to span a wide temporal range. The result is displayed up to $t=100 \mu$s because at this time the front has traveled over about $50$cm, which is comparable with the size of an experiment.
The initial droplet radius $R_0 = 25.4$nm is estimated using Eq.~\ref{eq.req} and is consistent with observations for similar initial temperatures \cite{Lescoute}. This size is the mean size of the liquid fragments so we are considering the evolution of a \textit{typical droplet}.

At early times ($t < 10$ns), a fraction of the liquid is evaporated (Fig.~\ref{fig.refcase}.a). However, this process saturates at a time $t_{\rm{min}}$, after which the liquid fraction starts \textit{growing} slowly. Then, in the purely 1D case (dashed lines), the droplet continues to grow steadily. In the more realistic situation however (solid lines), the droplet evaporates again when the 3D regime sets in, at times $t>t_{3D}$. 

In Fig.~\ref{fig.refcase}.b, one sees that, almost instantaneously after heating ($t < 100$ps), a temperature difference $\Delta T= T_l-T_g$ is established between the gas and the droplet, and remains roughly constant throughout the expansion in the 1D case. On the contrary, in the 3D case, $T_g$ drops quickly to almost 0 around $t_{3D}$, whereas $T_l$ decreases slowly to a value around 1600K. 

In the phase diagram trajectories [Fig.~\ref{fig.refcase}.d], we see that in both cases the liquid density remains very close to the equilibrium value. By contrast, the gas density is clearly below the binodal in the 1D case, and in the 3D case it dives deep into non-equilibrium (supersaturated) conditions.

In Fig.~\ref{fig.refcase}.c, we check the pressure equilibrium condition in the 3D expansion case. One cannot distinguish the pressure difference in the computed evolution (dashed lines) that uses Eq.~\ref{eq.Peq} from the theoretical value of Eq.~\ref{eq.Laplace} (dotted lines), as the agreement is better than 2\% over the whole simulation range. The increase of $P_l$ at long times is due to the Laplace term (Eq.~\ref{eq.Laplace}).

Clearly, from the NDCX II example, two different regimes can be identified. The first one, where the temperature difference is small, and remains constant, is a \textit{quasi-thermalized regime}. In this regime the droplet grows. The second one, where the gas becomes much colder than the drop, is a \textit{non-thermalized regime}. In this regime the droplet evaporates again, as if it were in vacuum.
We now discuss the various regimes.

\subsection{B. Thermalization condition, quasi-thermalized regime}
Let us find a  thermalization condition. In our equations, the energy is extracted from the system only by the adiabatic expansion of the gas (Eq.~\ref{eq.dEtot}) and the gas quenching is then transmitted to the liquid via the liquid-gas energy exchange term (Eq.~\ref{eq.dEl}).
Therefore, we should compare those two energy fluxes to find the thermalization condition.

Let us assume a small temperature difference $\Delta T/T  \ll 1$, so that we are in the \textit{quasi-thermalized} regime of expansion, as in the 1D case of Fig.~\ref{fig.refcase}. From Fig.~\ref{fig.refcase}.a, one sees that $N_l$ and $N_g$ are almost stationary if $T_l \simeq T_g$. Hence, let us make the approximation $\Phi_{\rm{vap}} = \Phi_{\rm{cond}}$ (more precisely, $\vert \Phi_{\rm{vap}} - \Phi_{\rm{cond}}\vert \ll  \Phi_{\rm{cond}}$).

The ratio between the two energy fluxes can be estimated as follows. Let us consider a cell containing fixed numbers $N_l$ and $N_g$ of liquid and vapor atoms, respectively, and let $x=N_l/N_{\rm{tot}}$ be the liquid fraction in the cell, where $N_{\rm{tot}}=N_l +N_g$. In the low $T$ limit, a small energy change can be written $d E_g= N_g \frac{3}{2} k_{B} d T_g$ for the gas and $d E_l= N_g \frac{5}{2} k_{B} d T_l$ for the liquid, according to Eq.~\ref{eq.energies}. Noting $d E_{\rm{tot}} = dE_l +dE_g$ the total energy lost by the cell, we define $\xi =d E_l /d E_{\rm{tot}} $. Requiring stationary $\Delta T$ (i.e. $d T_l = d T_g$), we obtain $\xi= 5 x/ (2 x +3)$.

In the low $T$ limit, the adiabatic cooling of the gas implies: $d E_{\rm{tot}}/dt = - P_g  d V/dt = -  n_g k_{B} T_g\tilde{ \eta} V_0$, where we define the volume strain rate $\tilde{\eta} = d(V/V_0)/dt$. Note that, in the 1D expansion regime, $\tilde{\eta} = \eta_z$, whereas in the 3D expansion regime, for $t \gg t_{3D}$, $\tilde{\eta} \simeq 3 \eta_z \eta_r^2 t^2$ and diverges.
The power transferred from the liquid to the gas is: $d E_l/dt = n_g \sqrt{ k_{B} T_g/2 \pi m} S \chi c_v \Delta T$,  where $\chi = \beta + (1-\beta) \alpha$. To get this expression we have computed the contributions from the three terms in Eq.~\ref{eq.dEl}  and used $\Phi_{\rm{vap}} = \Phi_{\rm{cond}}$.
Expressing $S= 4 \pi R^{2}$ and $V_0 = 2 \times \frac{4}{3} \pi R_0^3$, the balance between the fluxes $d E_l = \xi  \ d E_{\rm{tot}}$  finally yields
\begin{equation}
\frac{\Delta T}{T} = \frac{\xi\tilde{ \eta}}{\chi} \frac{4 \sqrt{2 \pi}}{9} \frac{R_0 ^{3}}{v_{\rm{th}}(T_g) R^2} 
\label{eq.deltaT}
\end{equation}
where we note $v_{\rm{th}}(T_g) = \sqrt{ k_{B} T_g/m}$ the thermal speed in the gas. 
With the approximation $R_0 \simeq R$ and neglecting the prefactors of order unity, we
obtain the simple scaling that is expected to be valid for any EoS :
\begin{equation}
\frac{\Delta T}{T} \sim   \frac{ \tilde{ \eta} \  R_0}{ v_{\rm{th}}(T_g)}.
\label{eq.deltaT2}
\end{equation}

\begin{figure}[htbp]
\begin{center} 
\includegraphics[width=8.5cm]{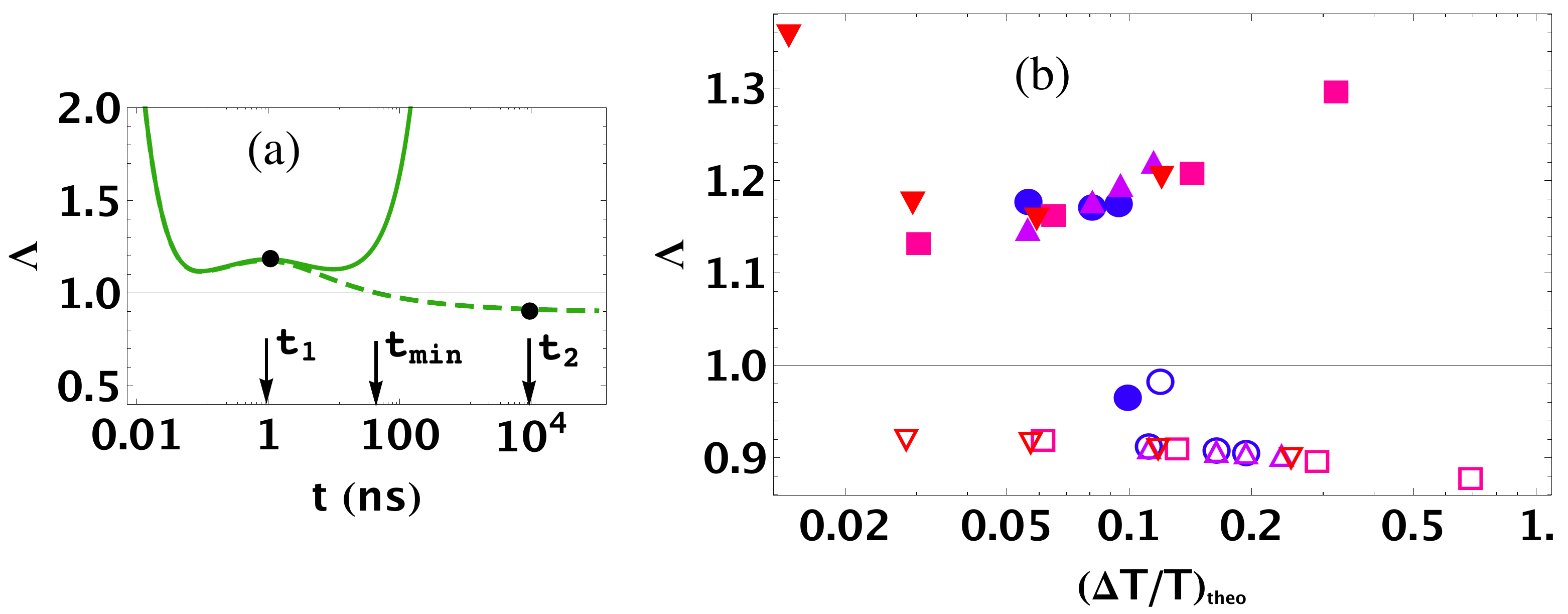}
\caption{(Color online)  Test of the thermalization formula Eq.~\ref{eq.deltaT}. (a) Time evolution of the ratio $ \Lambda = (\Delta T /T)_{\rm{theo}} /( \Delta T /T)$ of the temperature difference computed with Eq.~\ref{eq.deltaT} to the numerical model in the NDCXII reference case, in purely 1D (dashed line) and full 3D (solid line) expansion. (b) $\Lambda$ in the 1D expansion at $t_1=1$ns (solid symbols) and at $t_2=10\mu$s (open symbols) with $\beta=$0, 0.2, 0.4, 1 (circles), $\alpha =$0, 0.2, 0.4, 1 (up triangles), $\delta z=$1, 2, 4, 8$\mu$m (squares) and $R_0=$5, 10, 20, 40nm (down triangles).}
\label{fig.therm}
\end{center}
\end{figure}

In Figure ~\ref{fig.therm} we check the validity of Eq.~\ref{eq.deltaT}, computing the ratio $ \Lambda=(\Delta T /T)_{\rm{theo}}/( \Delta T /T)$ of the theoretical temperature difference (Eq.~\ref{eq.deltaT}) to the result of the full numerical calculation. To evaluate Eq.~\ref{eq.deltaT} we take the values of $\tilde{\eta}$, $R$, $T_g$ and $x$ from the result of the numerical simulation. In Fig.~\ref{fig.therm}.a, we show the evolution of the ratio $\Lambda$ in the NDCXII reference case (same calculation as Fig.~\ref{fig.refcase}). In the full 3D expansion case (solid line), the prediction becomes bad (error larger than 100\%) at $t\simeq t_{3D}$, which is expected since the volume expansion rate $\tilde{\eta}$ diverges in 3D. In the purely 1D expansion, after the first ns, one sees that Eq.~\ref{eq.deltaT} is accurate within 20\%. More precisely, the error has the same sign as the derivative $dN_l/dt$ and vanishes when the droplet is stationary, at $t=t_{\rm{min}}$, which is expected since Eq.~\ref{eq.deltaT} is obtained with the assumption $dN_l/dt=0$. 

In Fig.~\ref{fig.therm}.b, we show the ratio $\Lambda$ at $t_1=1$ns and at $t_2=1\mu$s for the same parameters, but varying one by one those that are relevant to Eq.~\ref{eq.deltaT}: $\beta$, $\alpha$, $\delta z$ (in order to vary $\eta$) and $R_0$. The analytic formula overestimates (underestimates) $\Delta T/T$ in all cases at $t_1$ ($t_2$), and for both cases the error is larger when the expected $(\Delta T/T)_{\rm{theo}}$ is larger, which is natural since Eq.~\ref{eq.deltaT} holds in the limit of small $\Delta T/T$. Interestingly, the point $\beta=0$ is separate from the others at both times, and is closer to 1, which is not surprizing since $\beta =0$ means no particle exchange, and this again confirms that the main source of inaccuracy of Eq.~\ref{eq.deltaT} is a non-zero value of $dN_l/dt$. The prediction could thus be refined if this effect was taken into account, for example using the analytical results of the next section. At early times, one can see also that the error is larger for droplets of radii smaller than 5nm. This effect is switched off if we set $\sigma=0$, confirming that it is caused by surface effects.

It is also possible to rewrite Eq.~\ref{eq.deltaT2} in terms of the initial conditions: considering $\eta\sim c^s_0 / \delta z$, one gets the very simple scaling law: $\Delta T/T  \sim R_0/ \delta z$
This last expression is only valid in the case of a 1D linear expansion where $\tilde{\eta}$ is constant. In this case, one sees using Eq.~\ref{eq.req} that $\Delta T/T$ is expected to decrease slowly when the initial sample size increases
\begin{equation}
R_0 \ \propto \ \eta^{-\frac{2}{3}}  \ \  \Rightarrow  \ \  \frac{\Delta T}{T}  \  \propto \   \eta^{\frac{1}{3}}   \  \propto  \    \delta z^{-\frac{1}{3}} .
\label{eq.deltaT0}
\end{equation}
For larger samples, the thermalization will be better even though the droplets are bigger. This justifies again that in the limit of large samples and slow expansions, an equilibrium hydrodynamic description becomes valid.

In summary, Eq.~\ref{eq.deltaT} is expected to be always a good estimate in the quasi-thermalized regime, and Eq.~\ref{eq.deltaT2} can be considered as a universal criterion to delimit the quasi-thermalized regime.

\subsection{C. Fully thermalized regime}

In the previous section we could distinguish the regimes of quasi-thermalized versus non-thermalized expansion. From Eq.~\ref{eq.deltaT} it is clear that the quasi-thermalized regime will become quickly invalid after $t_{3D}$, because the volume strain rate $\tilde{\eta}$ diverges.
Nonetheless, in the early times of expansion, or if one is interested in systems of large radial extent, it is worth studying the limiting case of a fully thermalized flow.

In this perspective, let us assume $T_l=T_g = T$.
Again, we look at the low $T$ regime, which becomes valid very early in the expansion process.
Using the first order approximations in the VdW model $n_l \simeq (1-\theta)/b$ and $n_g\simeq 0$ (Eq.~\ref{eq.nl} and ~\ref{eq.ng}) and neglecting the surface energy term, we write the total energy $E_{\rm{tot}} = N_{\rm{tot}} ( c_{v} T - x (1- \theta)a/b)$, where $x=N_l/N_{\rm{tot}}$ is still the liquid fraction in the cell. 
The total energy change $d E_{\rm{tot}} = - P_g dV$ becomes, at first order in $\theta$: $- (1-x)\  \theta \  \frac{dV}{V} = (\frac{3}{2} +x) \ d\theta - dx$. Noting that $dV\simeq dV_g$, we convert $ \theta\  d (\ln(V)) =  d \theta - d \theta / \theta$ using the low $T$ approximation Eq.~\ref{eq.ng}, and find finally
\begin{align}
dx =  \bigg(\frac{5}{2} - \frac{1-x}{\theta}\bigg) \ d\theta.
\label{eq.therm}
\end{align}
It is easy to push the approximation to higher orders in $\theta$, but Eq.~\ref{eq.therm} already allows one to get good insight into the evolution of the droplet. One sees that the droplet will be stationary at a temperature satisfying $\theta_{\rm{min}} = \frac{2}{5} (1-x)$ corresponding to the time $t_{\rm{min}}$ already mentioned, it will evaporate before this point, for temperatures $\theta > \theta_{\rm{min}}$, and grow after it, for $\theta < \theta_{\rm{min}}$. This sequence is in agreement with the NDCX II reference case shown in fig.~\ref{fig.refcase}. Note that, at long times, and independently from the EoS, droplets will always grow in a thermalized situation. This is due to the fact that adiabatic expansion of a perfect gas is an algebraic trajectory in phase-space ($T \propto \rho^{2/3}$), whereas Clausius-Clapeyron law predicts an exponential curve for the gas binodal, meaning that the gas in a two-phase expanding cell will always tend to saturate and make the liquid fraction grow.

\begin{figure}[htbp]
\begin{center}
\includegraphics[width=8.5 cm]{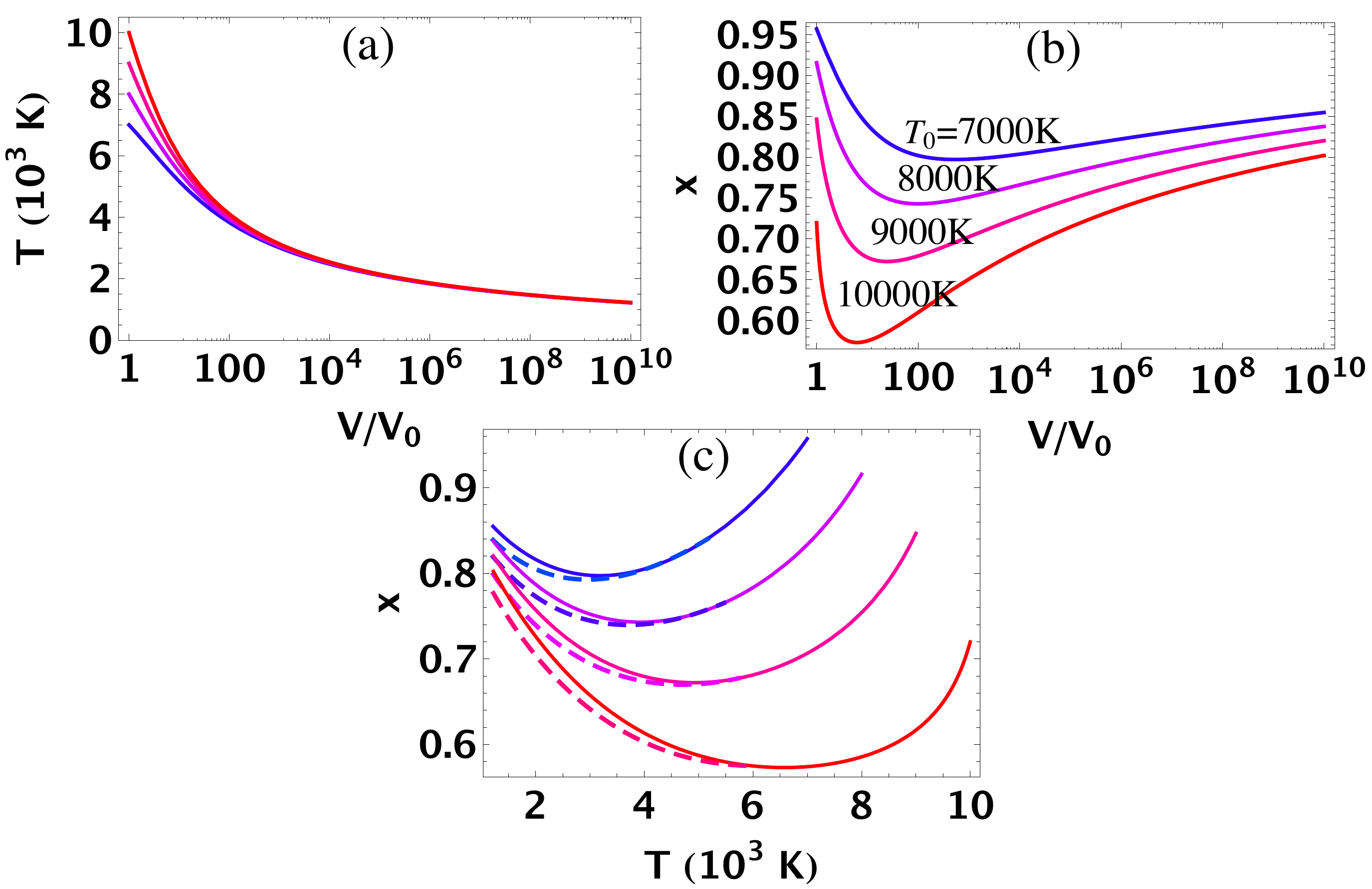}
\caption{(Color online)  Thermalized evolution for a droplet of initial radius $R_0=25$nm and initial temperatures $T_0=7000~\rm{K}~\rm{to}~10000 K$ (solid lines). Temperature (a) and liquid fraction (b) evolution versus the cell expansion. (c) Full numerical calculation compared to solution of Eq.~\ref{eq.therm} (dashed lines) started at $V/V_0 =10$.}
\label{fig.thermalized}
\end{center}
\end{figure}

In Figure~\ref{fig.thermalized}, we show the exact numerical calculation of the thermalized evolution of a droplet whose initial radius is $R_0 = 25$nm and for initial temperatures ranging from 7000 to 10000K. In the thermalized case, the time evolution is irrelevant, that is why in fig.~\ref{fig.thermalized}.a and b the temperature and liquid fraction are plotted as a functions of the volume expansion ratio $V/V_0$. An expansion remaining in the thermalized regime over 10 orders of magnitude volume expansion is unrealistic in the case of NDCX II, but for generality it is interesting to study this limiting trend. 

In Fig.~\ref{fig.thermalized}.c, the numerical liquid fraction versus temperature is compared to the solution of Eq.~\ref{eq.therm}, starting when the volume has expanded by one order of magnitude. The analytic approximation is accurate within 20\%.
This shows that Eq.~\ref{eq.therm} can be used to make good estimates of the asymptotic growth of the liquid fraction in the thermalized case, which can be, e.g. for $T_0= 9000$K, of a bit more than 10\%. 

This growth of the liquid fraction in the thermalized regime is a rigorous upper bound for droplet growth. In the opposite regime, one can get the reciprocal upper bound for droplet evaporation.

\subsection{D. Non-thermalized regime: evaporation in vacuum}

If the gas expands too fast for thermalization to occur, one expects the droplet to evaporate as if it were in vacuum. The corresponding limit consists of assuming $\Phi_{\rm{cond}}=0$. Obviously, in this case the droplet can only lose particles. Moreover, the evaporation of the droplet is maximal in this regime, because, if there was thermalization via collisions with a gas, colder than the drop, it would be a way for the droplet to lose energy without losing particles. 
Also, because the vapor pressure decreases exponentially with temperature, one expects the evaporation in vacuum to slow down fast. However, independently from the kinetics, it is clear that there must be some upper bound to the evaporation of a drop. Indeed, as every evaporating particle takes away energy from the droplet (the latent heat), the droplet gets colder and colder, until the evaporation is "frozen", a strict limit being that $T$ cannot become negative.

Let us find analytic expressions for the maximal evaporation of a droplet whose energy is noted $E_l$. Considering the evaporating particles and using Eq.~\ref{eq.dEl}, the energy loss can be written $dE_l = e_g^* dN_l $. On the other hand, considering the liquid, and neglecting the surface energy term, one has $dE_l = e_l dN_l + N_l (\partial{e_l}/\partial{T_l})dT_l$. Equating those two expressions, one finds, for the VdW fluid :
\begin{equation}
\frac{dN_l}{N_l}=\frac{(c_v - a( \partial{n_l}/\partial{T_l})) }{ a(n_l - n_g)} dT_l.
\label{eq.evap0}
\end{equation}
Using the development of $n_l$ at first order in $\theta$ (Eq.~\ref{eq.nl}), one can integrate Eq.~\ref{eq.evap0} from initial $T_0$ and $N_{l0}$, yielding at the final $T_l$ the remaining fraction
\begin{equation}
\frac{N_l}{N_{l0}}= \exp{ \bigg[ - \frac{5}{2} (\theta_0 -\theta_l)-\frac{13}{4} (\theta_0^2 - \theta_l^2) \bigg]} .
\label{eq.evap1}
\end{equation}

\begin{figure}[htbp]
\begin{center}
\includegraphics[width=8.5 cm]{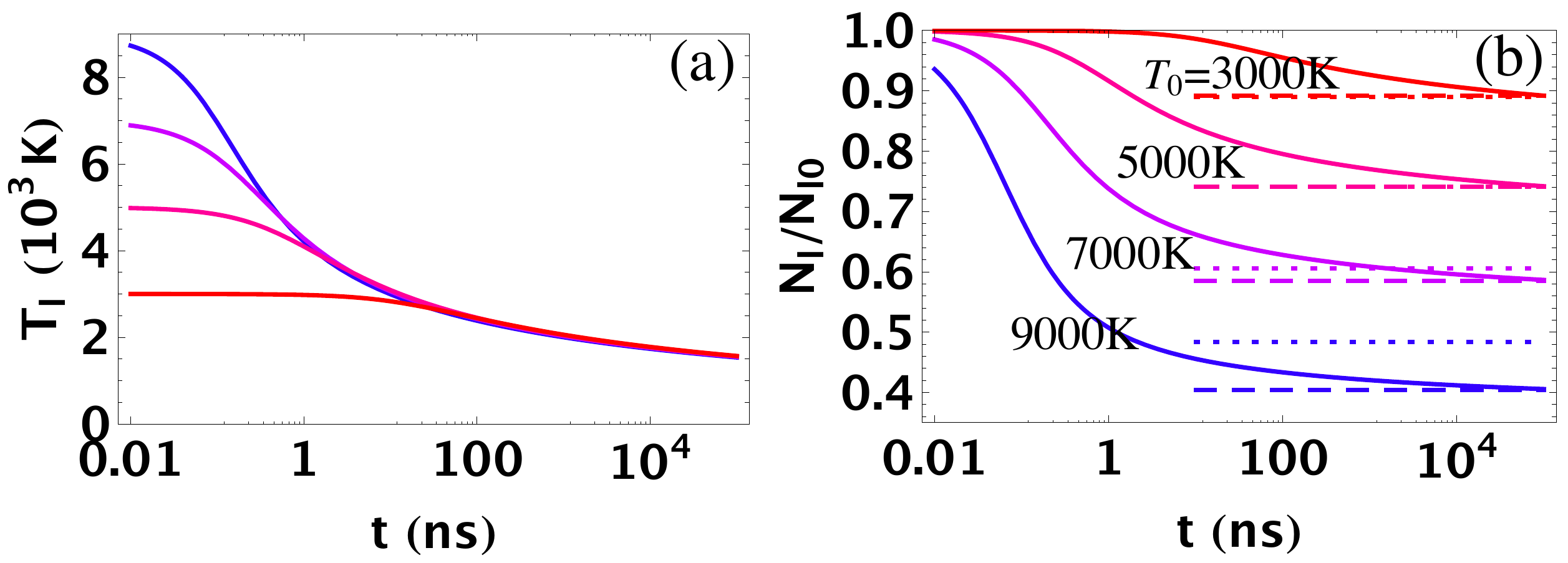}
\caption{(Color online)  Evaporation in vacuum. Time evolution of the temperature (a) and the non-evaporated fraction (b) for a droplet of initial radius $R_0 = 25$nm and initial temperatures $T_0 = 3000 \rm~{to}~ 9000$K (solid lines). The dashed lines on (b) are the numerical integration of Eq.~\ref{eq.evap0} and the dotted lines correspond to Eq.~\ref{eq.evap1}}. 
\label{fig.evapvac}
\end{center}
\end{figure}

Figure~\ref{fig.evapvac} shows the exact numerical result for the evaporation in vacuum of a droplet whose initial radius is 25nm, and initial temperatures ranging from 3000 to 9000K. The volume expansion is not relevant here, so the variables are displayed as a function of $t$ only.
In Fig.~\ref{fig.evapvac}.a and \ref{fig.evapvac}.b, one sees that for all initial temperatures, the number evaporation and cooling curves of the droplet follow a same asymptotic behavior, which is increasingly slow at long times. In Fig.~\ref{fig.evapvac}.b, the numerical integration of Eq.~\ref{eq.evap0} is shown for each $T_0$ (dashed lines), with the final $T_l$ taken from the full numerical solution. The agreement with the final evaporation ratio is excellent, showing that surface effects play a negligible role. We have checked that surface effects cause an overestimation of the maximal evaporation of less than 10\% even for droplets of initial radius 1nm. The approximated Eq.~\ref{eq.evap1} is also displayed for each $T_0$ (dotted lines), taking here also the final $T_l$ from the full numerical run. One sees that it predicts the good limit for evaporation within 5\% for initial temperatures up to 7000K. This is very satisfactory because the non-thermalized regime is expected to be valid only in the late times of expansion so the first order low $T$ approximations should be very accurate.

Let us now discuss the onset of the radiative cooling regime. At long times and low temperatures, the particle evaporation and the evaporative cooling rates decrease exponentially (Eq.~\ref{eq.ng}), whereas the radiative cooling rate is algebraic $(\propto T^{4})$. Therefore thermal radiation becomes the dominant cooling mechanism at long times.
Within our model it is not difficult to express the temperature $T_{\rm{rad}}$ below which radiative cooling becomes dominant over evaporative cooling \footnote {The evaporative energy flux per unit area is well approximated in the low $T$ limit by: $\Phi_E^{evap} \simeq \beta \Phi_{\rm{vap}} l_0 \simeq (\beta/b \theta) e^{-\frac{1}{\theta}} (k_{B} T/2 \pi m)^{1/2} l_0$, while the radiative energy flux is simply: $\Phi_E^{rad} = \epsilon \sigma_{S} T^4$ where $\epsilon$ is the emissivity of the liquid and $\sigma_{S}$ the Stefan constant. Finally the two processes are equally important at a temperature verifying $\theta^{-9/2} e^{-1/\theta} = (b \epsilon \sigma_{S}/\beta)  l_0^{5/2} k_{B}^{-3/2} (2 \pi m)^{1/2}$. To apply this formula to Si, we obtain the VdW parameters $a_{\rm{Si}}= 10.1 \times 10^{-35}$erg.cm$^{3}$ an $b_{\rm{Si}}=1.68 \times 10^{-23}$cm$^{3}$ by fitting the VdW EoS to the experimental values of the latent heat $l(T_b) = 359$kJ/mol at the boiling temperature $T_b=3530$K and of the liquid density $\rho_l(T_m)=2.57$g/cc at the melting point $T_m=1687$K \cite{CRC}, in the same way as was done for Al in Sec. III.A.}.
Using our VdW parameters for Al, and assuming an emissivity $\epsilon=0.2$, we find $T_{\rm{rad}} \simeq 1740$K. 
For Si nanoparticles, the same estimate yields $T_{\rm{rad}} \simeq 2190$K. This is consistent with the measurements reported in \cite{AmorusoSi} where the cooling of Si nanoparticles formed by laser ablation is found to be well explained by radiation at expansion times $5 - 150 \mu s$ and for temperatures below 2000K, although the evaporative cooling rates that we obtain within our model are significantly larger than the estimates they report. As an example, for Si at $2000$K, our crude model predicts a radiative cooling rate of $\simeq$28K/$\mu$s, while the evaporative coling rate in vacuum is still $\simeq$6.5K/$\mu$s. Note however that the rate we compute is a strict upper bound.

\subsection{E. Supercritical case: nucleation and growth of liquid droplets}

The last case to consider is the supercritical case, where the material expands first as a supercritical fluid, and enters the two-phase region of the phase diagram crossing the gas binodal, as a supersaturated gas. In this case nucleation of droplets may occur. 

We do not propose a model for nucleation but we note that it has been reported \cite{Lescoute} that supersaturation of the vapor does not reach high values and that above a certain threshold value nucleation is very sudden, due to the exponential dependence of the nucleation frequency on the supersaturation ratio \cite{Balibar}. Then, our model is expected to describe correctly the subsequent evolution of the clusters. In particular, we expect that thermalization will depend on the droplet radius and the volume strain rate and that Eq.~\ref{eq.deltaT} will be a good estimate in the quasi-thermalized regime. If clusters and gas are in thermal equilibrium, we expect Eq.~\ref{eq.therm} to be valid as well.

Note that nucleation may also happen in a subcritical expansion scenario, if the gas becomes very supersaturated at long times, as can be seen for the reference case in Fig.~\ref{fig.refcase}.d. In this situation, close to the non-thermalized limit, condensation on the existing droplets is too slow, and nucleation of new droplets may happen even with liquid clusters already present in the plume.

\section{V. Conclusion}

In conclusion, we have studied droplet evolution and thermalization conditions with an original, simple kinetic model based on a consistent set of rate equations for mass and energy exchanges. Using the vdW EoS as a test-bed, we have demonstrated that such a kinetic treatment is able to 
bridge the gap between the molecular and equilibrium hydrodynamic approaches that have mainly been used so far. Most of the results of this work are general and should be extendable to any EoS with which the kinetic equations are used.

In particular, the main output of our study is to identify the different regimes of two-phase expansion. On one side, the quasi-thermalized case and its limit, the fully-thermalized case, on the other side, the non-thermalized case. To distinguish the two situations, we identify a local thermalization condition (Eq.~\ref{eq.deltaT}) which depends on the droplet radius $R$, the volume expansion rate $\tilde{\eta}$, the gas temperature $T_g$, the liquid fraction $x$, and the kinetic parameters $\alpha$ and $\beta$, but it can also be traced back to the initial conditions: sample thickness $\delta z$ and initial temperature $T_0$ (Eq.~\ref{eq.deltaT0}). Eq.~\ref{eq.deltaT2} is a simpler alternative to Eq.~\ref{eq.deltaT} that involves only the initial droplet radius $R_0$, $\tilde{\eta}$ and $T_g$ and gives the scaling expected for any EoS.

Due to the crossover from 1D to 3D expansion at the time $t_{3D}$, the expansion is expected to take place in the quasi-thermalized regime at early times, at least in the NDCXII reference case and for similar parameters, but at long times the non-thermalized regime is almost inevitable. Eq.~\ref{eq.deltaT} shows that this is only a dimensional effect, driven by the divergence of $\tilde{\eta}$ in 3D.

In the quasi-thermalized case, our study shows that the relative temperature difference $(T_l-T_g) /T_l$ remains almost constant throughout the expansion. Eq.~\ref{eq.deltaT} is derived assuming no net particle exchange, so only kinetic energy terms (but no latent heat) are involved, which makes it suitable for generalization to other EoS.
The predictions of Eq.~\ref{eq.deltaT}, and, more generally, the validity of the whole droplet evolution model, could also be tested with MD calculations and experimental measurements.

In the fully thermalized case, droplets can grow (moderately) at long times and we give an approximate formula for the VdW fluid (Eq.~\ref{eq.therm}).
In the opposite case of a fully non-thermalized flow, one can find a strict upper bound for the evaporated fraction at a given final $T_l$. For the VdW fluid, this limit is Eq.~\ref{eq.evap0}.
In both limiting cases, simple implementations of the kinetic  model, e.g., with the VdW EoS, can be used to make estimates and provide upper bounds for the droplets evolution.
Note that, in all the cases we have studied, the droplets never grow or evaporate very much from their initial situation.

For the moment, the model we have presented is local, but in the future it could become part of a larger hydrodynamic code that will treat many lagrangian two-phase cells containing droplets and gas. The extension of our model to several droplets in one cell can be done easily. 
For more realistic simulations, it will probably be necessary to take into account other effects that are not two-phase phenomena and that we have left aside, such as radiation, which can be non-blackbody in the early stages of expansion, but also, thermal conduction between cells, and thermionic emission. 

Before our kinetic model can be used to compute droplets evolutions at the core of a global comprehensive code, additional modules are needed to initialize the two-phase regime.
In the subcritical case, a hydrodynamic code and some model for fragmentation is required to determine the droplets distribution at each location, and possibly their velocities, which may differ from that of the expanding gas.
In the supercritical case, a model for nucleation is needed. In any situation, the thermalization condition (Eq.~\ref{eq.deltaT} or \ref{eq.deltaT2}) can be used to check wether the two-phase computing cell can be treated as an equilibrium cell or if a non-equilibrium treatment is required. The reason being of course that an equilibrium (thermalized) description is much easier to implement.

Finaly, we have investigated the role of surface effects in different cases.
Surface tension is expected to play an important role for droplets of radii $R < R_{\sigma}= \sigma / k_B T n_l$. This can be seen from Kelvin equation or considering the radius at which the surface energy becomes comparable to the kinetic energy per particle in the liquid. With our VdW parameters for aluminum, $R_{\sigma}$ increases from $0.8$nm at $T=10000$K to $64$nm at $T=2000$K. Surface effects are thus increasingly important in the late stages of expansion, at low temperature and for the smallest fragments. This is also why a careful treatment of the supercritical case of in-flight nucleation is more difficult and remains to be done in order to complement this work.

\section{Aknowledgements}

We wish to thank R. M. More, F. M. Bieniosek, P. A. Seidl, B. G. Logan and I. D. Kaganovich for helpful discussions. One of us (JA) was partially supported by Ecole Normale Sup\'erieure, France. Work performed under the auspices of the U.S. Department of Energy under University of California contract DE-AC02-05CH11231 at Lawrence Berkeley National Laboratory and contract DE-AC52-07NA27344 at Lawrence Livermore National Laboratory.

\bibliographystyle{prsty}
 \bibliography{droplets}

\end{document}